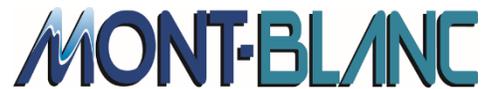

# The Mont-Blanc Project: First Phase Successfully Finished

Momme Allalen, David Brayford, Daniele Tafani, Volker Weinberg
Leibniz Supercomputing Centre (LRZ), Germany
contact: Momme.Allalen@lrz.de, David.Brayford@lrz.de, Daniele.Tafani@lrz.de, Volker.Weinberg@lrz.de

Bernd Mohr, Dirk Brömmel, Rene Halver, Jan Meinke, Sandipan Mohanty
Jülich Supercomputing Centre (JSC), Germany
contact: B.Mohr@fz-juelich.de, D.Broemmel@fz-juelich.de, R.Halver@fz-juelich.de, J.Meinke@fz-juelich.de, S.Mohanty@fz-juelich.de

With the ever-increasing energy demands and prices and the need for uninterrupted services, data centre operators must find solutions to increase energy efficiency and reduce costs. Running from October 2011 to June 2015, the aim of the European project Mont-Blanc [1] has been to address this issue by developing an approach to Exascale computing based on embedded power-efficient technology. The main goals of the project were to i) build an HPC prototype using currently available energy-efficient embedded technology, ii) design a Next Generation system to overcome the limitations of the built prototype and iii) port a set of representative Exascale applications to the system. This project was coordinated by the Barcelona Supercomputing Centre and had a budget of over 14 million EUR, including over 8 million EUR funded by the European Commission.

Improving the energy efficiency of future supercomputers is one of LRZ's main research goals. Hardware prototyping of novel architectures has proven to be successful for the technology watch preceding the selection of large supercomputers. Yet, advances in hardware are only justified if the need for programmability and thus the productivity of application development is still satisfied. Therefore, LRZ's contribution to the Mont-Blanc project was twofold: on the software side, application experts successfully ported two applications to the new system and analysed their performance and productivity. On the hardware side, LRZ's computer architecture experts worked on a system monitoring solution for acquisition and storage of Mont-Blanc sensor data, with particular emphasis on node energy consumption.

Contributions from JSC were also twofold: teams from the Jülich Simulation Laboratories ported their applications to the new system and to the OmpSs programming model [2], and the cross-sectional team Performance analysis ported their well-known performance analysis tool components Score-P [3] and Scalasca [4] to the ARM and ARM64 architecture and adapted them to work with the OmpSs programming model [1, Deliverable D5.9].

One of the highlights of the project was the installation of the Mont-Blanc system at the Barcelona Supercomputing Centre. Fig. 1 shows the final prototype, consisting of 2 racks, each of which containing 4 standard BullX chassis. Every chassis fits 9 blades, hosting 15 compute nodes each. Every Mont-Blanc compute node comes with a Samsung Exynos 5 Dual System on a Chip, which includes an ARM Cortex-A15 @ 1.7 GHz dual core CPU and an ARM Mali T-604 GPU. In total, the Mont-Blanc system offers 2160 ARM CPU cores and 1080 ARM GPUs, making it a "first-of-a-kind" cluster of this size based on ARM architecture.

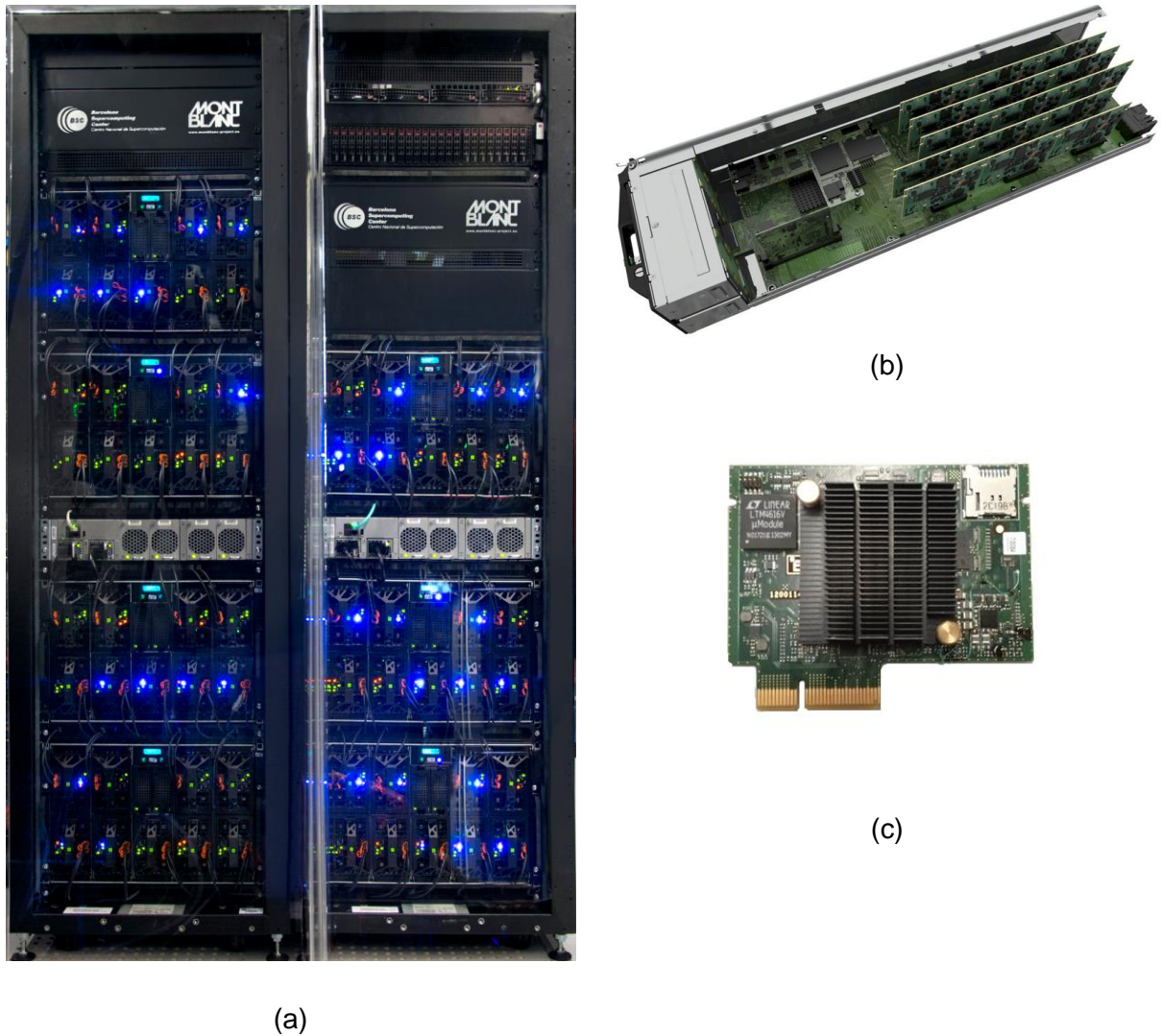

(b)

(c)

(a)

*Figure 1: The new Mont-Blanc prototype system at BSC, Barcelona (a), the Mont-Blanc blade (b) and the Mont-Blanc compute card (c).*

The performance of the system was thoroughly tested and evaluated with many real-life application codes: LRZ was involved with two different codes: the lattice Quantum Chromodynamics code BQCD as a representative of a real-world application and the Himeno benchmark. Berlin Quantum Chromodynamics (BQCD) [5] is a Hybrid Monte-Carlo program for simulating lattice QCD with dynamical Wilson fermions. The most important part of the program is a standard conjugate gradient solver. The Himeno benchmark [6] is the kernel of an incompressible fluid analysis code and focuses on the solution of a 3D Poisson equation. It is highly memory intensive and bound by the memory bandwidth. Both codes were successfully ported to the ARM architecture and tested on the Mont-Blanc prototype system. Beyond evaluating the original MPI-only and hybrid MPI+OpenMP versions on the prototype, LRZ also successfully ported BQCD to the OmpSs data flow programming model developed at BSC and investigated the scaling behaviour of code versions combining OmpSs with OpenCL and/or MPI.

JSC ported, tested and evaluated four applications codes: MP2C [7] implements the multi-particle collision dynamics method, which is a particle based description of hydrodynamics taking into account thermal fluctuations and making it possible to simulate flow phenomena on a mesoscopic level. PEPC [8] is a tree code for solving the N-body problem. It is not restricted to Coulomb systems but also handles gravitation and hydrodynamics using the vortex method as well as smooth particle hydrodynamics (SPH). PEPC is a non-recursive version of the Barnes-Hut algorithm with a level-by-level approach to both tree construction and traversals. PROFASI [9] is a Monte Carlo simulation package for protein folding and aggregation simulations. It implements an all-atom protein model, an implicit solvent interaction potential and several modern Monte Carlo methods for simulation of systems with rough energy landscapes. Finally, SMMP [10] provides advanced Monte Carlo algorithms and several force fields to simulate the thermodynamics of single proteins and assemblies of peptides. To port them to the Mont-Blanc prototype, MPI+OmpSs or MPI+OpenCL versions of the codes were developed and evaluated. Fig. 2 shows the power profile of a benchmark run of SMMP.

More details on the conducted performance analysis by LRZ and JSC can be found in [1, Deliverable D4.4].

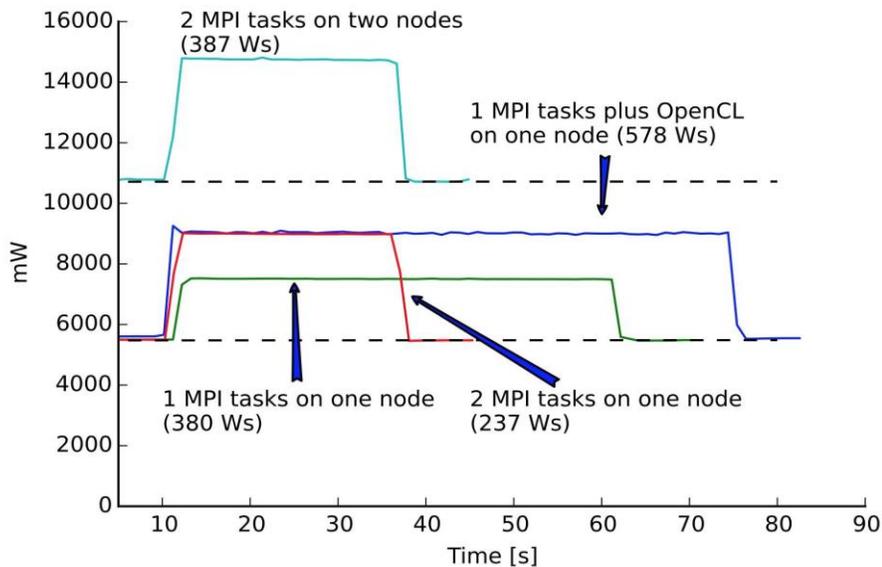

*Figure 2: Power profile of SMMP. The run with OpenCL uses the GPU in addition to the CPU. The numbers in parenthesis are the energies to solution (1 Ws = 1 J).*

Performance is the most important goal of HPC. However, programming languages should not be judged by the performance that can be reached alone, but also by the ease-of-use, i.e. the programmability. The combination of performance and programmability is commonly referred to as "productivity". To assess the productivity, various software engineering metrics like lines of code and time to solution can be used. For the productivity analysis within the Mont-Blanc project LRZ's software experts concentrated on the number of source lines of code to assess the ease-of-use of parallel programming languages. The number of source lines gives a rough estimate of the time necessary to program the code, as well as of the readability and maintainability of the code. The productivity analysis has been performed for various code versions of BQCD and the Himeno benchmark using several combinations of parallel programming languages like OpenMP, MPI, OpenACC, CUDA, OpenCL and OmpSs. To show an example of the productivity analysis, Fig. 3 presents the number of total lines of codes for different versions of the Himeno benchmark. Further productivity related results are discussed in [1, Deliverable D3.6].

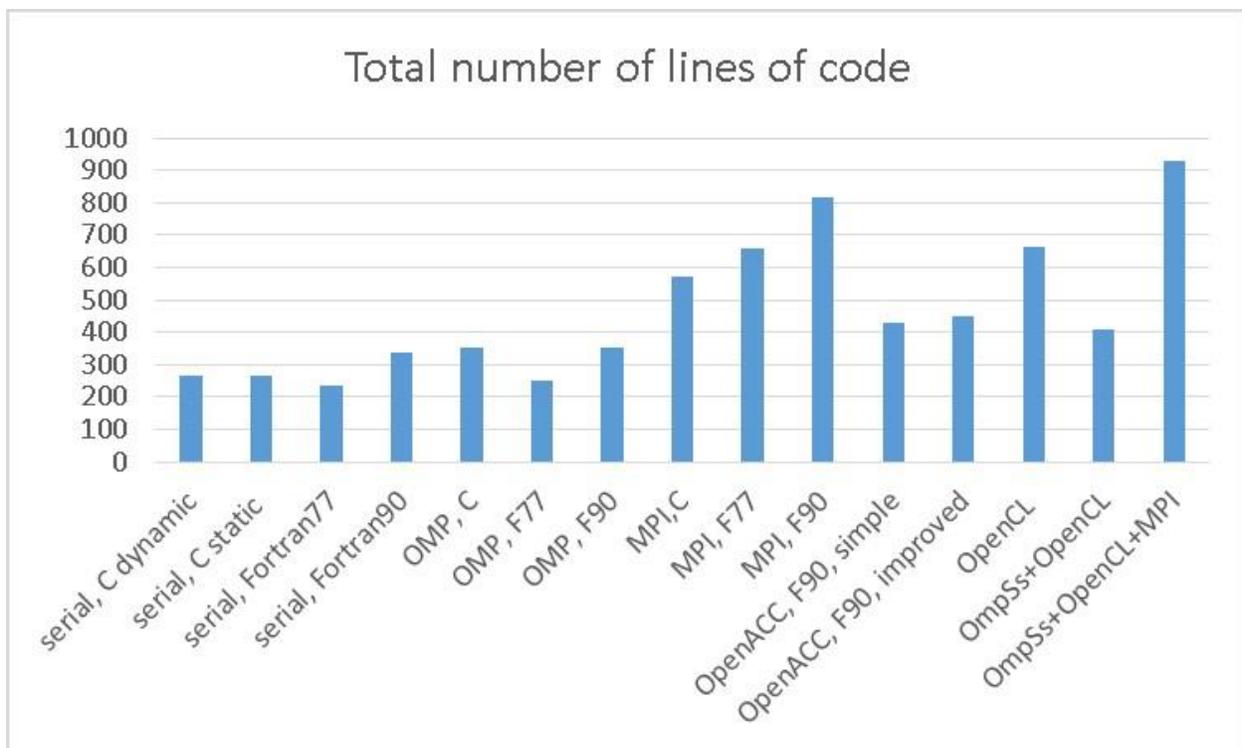

*Figure 3: Number of total lines of code for various versions of the Himeno benchmark.*

Finally, one of the key requirements for achieving energy efficiency is the capability of retrieving and storing detailed information on the power consumption of the HPC system. Working in close collaboration with Bull, LRZ researchers developed a holistic monitoring solution for the Mont-Blanc prototype, allowing not only to keep track of the power consumed by the system, but also and especially exposing this information to the system user. The monitoring tool was customised to cope with the characteristics of the Mont-Blanc system and features a low overhead transport messaging protocol and a scalable database for storing the monitored data [1, Deliverable D5.8]. In addition to standard cluster monitoring features and following the integration with the SLURM workload management system, the tool can also be used for implementing energy-aware job scheduling. Additionally, this integration will offer the opportunity of defining strategies for energy-aware user accounting. These last research topics alongside with the provision of performance and debugging analysis tools will be further addressed respectively by LRZ and JSC within the second phase of the project, running from 2013 until 2016. The second phase will complement the efforts of the first phase by targeting the system programmability, performance analysis and resiliency. In this phase we will also monitor the evolution of upcoming ARM-based devices and we will define the next Mont-Blanc Exascale architecture, investigating hardware design alternatives and their implications to the current system.

## *Acknowledgements*

The research leading to these results has received funding from the European Community's Seventh Framework Programme (FP7/2007-2013) under the Mont-Blanc project [1], grant agreement n° 288777 and n° 610402. We would like to thank Dr. Hinnerk Stüben (University of Hamburg) for his continuous collaboration.